# Metacognitive threshold: a computational account

**Brendan Conway-Smith  (brendan.conwaysmith@carleton.ca),**
**Robert L. West (robert.west@carleton.ca)**
Department of Cognitive Science, Carleton University, Ottawa, ON K1S 5B6 Canada

**Abstract**

This paper will explore ways of computationally accounting for the metacognitive threshold — the minimum amount of stimulus needed for a mental state to be perceived — and discuss potential cognitive mechanisms by which this threshold can be influenced through metacognitive training and meditation.

## Introduction

The ultimate goal of Cognitive Modeling is to build a Unified Cognitive Architecture that can simulate most, if not all, human cognitive abilities (Newell, 1994). Cognitive architectures like ACT-R and SOAR (Anderson & Lebiere, 1998; Laird, 2012) have achieved notable success in modeling knowledge-driven behaviour, however there is a scarcity of models related to phenomena surrounding metacognition. The ability for cognition to monitor and control its own processes, "metacognition," has risen to the forefront of research in psychology, psychiatry, and AI. Modeling the results of empirical studies of metacognition is important for making progress toward accurately describing human cognition.

This paper will address a cognitive phenomenon referred to as the metacognitive threshold, i.e., the minimum level of stimulus needed for a mental state to be perceived. Specifically, we will address the variability of the metacognitive threshold, which can be reliably lowered to allow an agent improved perceptual access to their own internal cognitive states (Pauen & Haynes, 2021). The degree of an individual's introspective acuity is also referred to as "metacognitive sensitivity". This can be reliably improved and the metacognitive threshold lowered by way of metacognitive training such as employing mindfulness techniques (Fox et al., 2016). In cognitive psychology, mindfulness is defined as deliberate attention directed toward perceptible mental experiences, i.e., affect, sensations, thoughts, etc. (Holas & Jankowski, 2013). Greater access to and control of one's own mental states have shown to strongly correlate with improved psychological health and overall cognitive functioning (Grossman et al., 2004; Tang et al., 2015; Rigby et al., 2014).

While decades of research strongly support the effectiveness of metacognitive techniques to influence one's metacognitive threshold, the underlying cognitive mechanisms have remained poorly understood. Presently there exists little or no account of this phenomena — the cognitive and computational underpinnings by which the metacognitive threshold is raised or lowered.

This paper will investigate potential computational mechanisms that may contribute to the lowering of the metacognitive threshold. In particular, we will discuss metacognitive techniques that have shown to increase metacognitive sensitivity, and explore various frameworks for clarifying their underlying cognitive constituents.

For this purpose, we will employ the Common Model of Cognition (CMC), originally the 'Standard Model' (Laird, Lebiere, & Rosenbloom, 2017) which provides a unified framework for investigating the fundamental elements of cognitive and metacognitive phenomena. By utilizing the Common Model and specifically ACT-R in this endeavor, we intend to address unanswered questions regarding the architecture and particularly concerning the nature of production rules.

## Metacognition

Metacognition refers to the monitoring and control of cognitive processes (Flavell 1979; Fleming, Dolan, & Frith, 2012). It also involves a wide range of introspective attitudes such as confidence ratings and judgments of learning (Frazier, Schwartz, & Metcalfe, 2021; Rhodes, 2016).

*Metacognitive control* involves the active regulation of cognitive states or processes (Proust, 2013; Wells, 2019). This involves engaging in mental actions to either access or suppress cognitive states. Mental actions are distinct from world-oriented actions. The control of cognitive activity can involve a range of processes such as attention, emotion, planning, reasoning, and memory (Slagter et al., 2011; Efklides, Schwartz, & Brown, 2017; Pearman et al., 2020).

*Metacognitive monitoring* refers to the ability to recognize and identify cognitive states. It involves the perception of internal mental states such as thoughts and feelings in order to regulate those states or direct behavior. Research has demonstrated that metacognitive monitoring can be developed and improved through training (Baird, Mrazek, Phillips, & Schooler, 2014). For instance, attentional processes can be developed and enhanced through the repeated practice of attention-based tasks (Posner et al., 2015). In particular, mindfulness as a form of attention has shown to develop through the three stages of skill acquisition defined by Fitts and Posner (Kee, 2019).

Metacognitive training such as mindfulness techniques plays a significant role in the success rates of Cognitive Behavior Therapy (CBT) and Metacognitive Therapy (MCT). Both CBT and MCT instruct patients on metacognitive strategies to monitor and regulate their own thoughts and emotions (Dobson, 2013; Normann & Morina, 2018). Research has demonstrated that those with improved metacognitive skills are better equipped to identify and manage their own disruptive and harmful thoughts and emotions (Wells, 2011, 2019; Hagen et al., 2017).

## Metacognitive monitoring as mindfulness

Metacognitive monitoring and mindfulness are often used interchangeably within cognitive psychology (Holas & Jankowski, 2013). Scientific interest in mindfulness practice has become a target of interdisciplinary research and has grown exponentially over the past few decades (Tang, 2017; Van Dam et al., 2018).

Mindfulness involves the deliberate focus on perceptible experiences (sensory, affective, thought-related) and the cultivation of a dispassionate awareness of mental states and processes (Brown & Ryan, 2003; Grossman, 2010). Studies indicate a technique called detached mindfulness to be a uniquely effective therapeutic practice in developing adaptive monitoring and control over maladaptive cognitive processes (Wells & Matthews, 1994; Wells, 2005).

Detached mindfulness is characterized by the awareness of internal states (thoughts and emotions) without reacting to them — without trying to maintain or suppress them. This is achieved by way of attempting to perceive the momentary changes in mental events (such as the minute fluctuations of emotions) and letting them pass without emotional response. Mindfulness psychology contends that a significant degree of emotional distress and pathological symptoms are caused by the illusory perception of affective experience being more permanent than it actually is. This illusory perception is explained as the result of a high metacognitive threshold (poor metacognitive sensitivity) that does not allow for the subtle detection of affective fluctuations. Training in detached mindfulness aims to improve metacognitive sensitivity and one's perception of affective impermanence, also referred to as equanimity. In mindfulness therapies that do not promote equanimity, awareness alone may not be sufficient to increase subjects' psychological well-being (Cardaciotto et al., 2008). The increased, more skillful, capacity to perceive the impermanence of affective experience is considered a key mechanism responsible for decreasing emotional reactivity (Tang et al., 2015).

## Metacognitive threshold

A psychophysical threshold is the minimum amount of physical stimulus needed to evoke a perceptual response in a person (Rouder & Morey, 2009). Psychophysical thresholds and their variability have been researched in domains such as sound, vision, interoception, and others (Kingdom & Prins, 2009). In metacognition research, psychophysical thresholds have been studied in reference to the minimal level of a stimulus required for a person to be aware of some mental state and make a judgment about it (Charles, Chardin, & Haggard, 2020; Sherman, Seth, & Barrett, 2018; Pauen & Haynes, 2021). These include confidence ratings as well as the subtle fluctuations of affective experience.

Generally, it is believed that an individual's metacognitive threshold is variable and can be lowered by way of training attention to perceive the momentary variations of internal cognitive states (equanimity). The training of equanimity through detached mindfulness and meditation practice has shown to be effective at lowering one's metacognitive threshold and enhancing metacognitive sensitivity.

**Metacognitive sensitivity** is the extent to which one is able to perceive their own mental processes or states, including thoughts, feelings, and emotions (Fleming & Lau, 2014). Mindfulness training can increase metacognitive sensitivity, allowing one to better perceive the nuances of their own feelings and thoughts.

Various metacognitive strategies and meditation techniques can allow one to practice and improve certain cognitive processes. Meditation is an umbrella term for techniques that employ deliberate focus and engage neurocognitive processes that result in advantageous effects on brain and behavior (Fox et al., 2016).

Various meditation techniques have the reported effect of enhancing metacognitive sensitivity, enabling one to perceive a weaker signal strength from internal cognitive states. In the case of developing equanimity, one becomes more capable of detecting subtle variations within emotional stimuli, such as the rapid arising and passing of feelings, thoughts, and emotions.

**Meditation** can involve a variety of practices. We will use Vipassana meditation as an example. Vipassana meditation (in the tradition of S.N. Goenka) is an old and popular technique that largely focuses on cultivating equanimity — a refined perception and sensitivity to the momentary impermanence of affect and sensations. Regular practice of this technique has shown to result in various cognitive advantages, such as improving executive functioning, enhancing response inhibition, and control over automatic reactions (Chambers, Lo, & Allen, 2008; Andreu et al., 2019).

The Vipassana method engages practitioners in guided meditation that directs them to maintain attention on the impermanence of their own sensations (Kakumanu et al., 2018). During this process, practitioners monitor their affective and bodily sensations moment-to-moment, without evaluation or emotional reactivity.

Following this technique, practitioners report being able to detect increasingly subtle properties of their own mental states, including improved perceptual access to fluctuations in affect that were previously inaccessible. In other words, subjects report greater metacognitive sensitivity and a concomitant lowering of their metacognitive threshold.

## Modelling the phenomena

A computational model that accounts for the phenomena surrounding the metacognitive threshold must necessarily ask questions about the fundamental nature of the architecture. Which computational components might allow one to perceive subtler properties in internal signals? Does it require us to change the way we think about certain elementary units of the cognitive architecture? We discuss these questions with specific reference to ACT-R, however, the application is intended more generally to the CMC family of architectures (Note, because the focus is on ACT-R, references are made to production systems. Other CMC architectures, such as SOAR, use more complex mechanisms, but the issues raised here remain relevant).

The ACT-R cognitive architecture fundamentally distinguishes between procedural and declarative knowledge to explain the underlying components of skill, which accords with the literature in philosophy and psychology (Squire, 1992; Christensen, Sutton, & McIlwain, 2016). Declarative knowledge is formatted propositionally and structured within semantic networks. Procedural knowledge is commonly referred to by researchers as containing "procedural representations" (Anderson, 1982; Pavese, 2019). In Anderson's ACT-R model, procedural representations are computationally specified as "production rules" which are a dominant form of representation within accounts of skill (Newell, 1994; Taatgen & Lee, 2003; Anderson et al., 2019). Production rules, or "productions", transform information and change the state of the system to complete a task or resolve a problem. A production rule is modeled after a computer program instruction in the form of a "condition-action" pairing. Essentially, a production rule is a "pattern-directed invocation of action" (Stocco et al., 2021). It specifies a condition that, when met, performs a prescribed action. A production is also thought of as an "if-then" rule. *If* the condition is satisfied, *then* it fires an action. Production rules are considered to be central to human intelligence and fundamental to the realization of cognitive skills (Anderson, 1993). Neurologically, production rules are associated with the 50ms decision timing in the basal ganglia (Stocco, 2018).

## Modelling the metacognitive threshold

How might production rules account for an enhanced ability to detect internal cognitive signals and their variations? This could be accomplished through increasing the speed of production rules. Essentially, making productions faster, particularly productions that notice internal states, would increase the chances of picking up fleeting or intermittent signals related to emotions and noetic (epistemic) feelings, such as confidence and feelings of knowing (FoK).

A complete model of this phenomena would involve modelling internal signals, how they are detected, how they break into one's current awareness, and how metacognitive training can improve this. Here, we discuss only how production rule acceleration can occur (however, see West and Conway-Smith [2019] for an account of how affect and noetic feelings can be incorporated into this type of model).

With regard to the speed-up of production rules, there are at least four different mechanisms that could accomplish this:

### *1. The ticking clock mechanism*

Production rules fire when a fixed amount of time is up. The use of this mechanism produces production timing that is analogous to the intervals of a ticking clock. The timing for production firing is generally estimated to be 50ms. During this interval, productions that match the buffer conditions are identified. When this time is up, the matching production with the highest utility will fire. The timing of this process is based on neural functions that are generally considered to occur within the Basal Ganglia. Using this mechanism, production time could be sped up by shortening the clock speed. This could possibly occur through top-down feedback related to attention, as its influence has been observed in other psychophysical thresholds, such as improving perceptual sensitivity (MacLean et al., 2010).

### *2. The fire when ready mechanism*

Production rules fire when they are ready. ACT-R is essentially a fire-when-ready model. ACT-R assumes that it takes 50ms for a production to fire, but if no production rule matches the buffer conditions, ACT-R will wait until the buffer conditions change. For example, for memory retrieval, ACT-R waits for the knowledge chunk to be delivered into the declarative memory buffer and then fires the matching production. Hence, the overall time taken is the memory retrieval time plus 50ms. However, if an alternative production matches the buffer conditions before this occurs, then it will fire instead. Using this mechanism, production firing can be made faster by using productions that do not wait for information from memory or perception.

These types of productions can be generated through the production compilation mechanism in ACT-R.

3. *The narrow focus mechanism*

ACT-R is capable of multitasking and even mind wandering, if the appropriate productions are available. The simplest way of producing a faster rate of firing for a specific type of production is to maintain the buffer conditions such that only this type of production can fire. Under these conditions, ACT-R can be said to model a narrow focus of attention.

4. *The faster production mechanism*

Some productions may be faster than others. Productions range in the complexity of their internal actions (Taatgen, 2013). The consequences of this at the neural level could imply that more complex productions take longer than simpler productions. Stewart et al.'s (2010) neural model of the Basal Ganglia estimates that this would produce a range between approximately 34ms-44ms for simple productions, and 59-73ms for complex productions. If this is the case, then the use of simpler productions would speed up the firing time.

All of these mechanisms could lower the metacognitive threshold by speeding up productions and thus increasing fidelity. To be clear, this viewpoint does not require that a threshold exists *in fact*, only that the resulting effect would appear to be so. Hence, from an architectural standpoint, this particular issue is not regarding thresholds but rather the complexity and timing of production rules. We propose that increasing the rate of production rule firing can potentially account for reports of increased metacognitive sensitivity as a result of metacognitive training, and that Common Model type architectures can model this.

## Metacognitive proceduralization

The process by which simpler, faster production rules are developed through metacognitive training can be largely explained by way of metacognitive proceduralization. Proceduralization is a concept used in the skill acquisition literature to explain the cognitive mechanisms involved. It refers to the process by which a task or skill becomes automated, allowing it to be performed more efficiently and accurately, with minimal conscious effort or attention. The process involves the converting of slow declarative knowledge into fast procedural knowledge that is increasingly refined. Performance can be further improved by mechanisms such as time delayed learning, where faster productions are rewarded.

Proceduralization plays a significant role in the cognitive processes underlying skill learning in domains such as motor skill and cognitive skill (Ford, Hodges, & Williams, 2005; Beilock & Carr, 2001; Anderson, 1982; Tenison & Anderson, 2016).

Conway-Smith, West, and Mylopoulos (2023) propose that metacognitive skill develops largely through the process of proceduralization. Based on the skill acquisition model of Fitts (1964) and Anderson (1982), this model relies on the principle that skill learning within any domain is principally realized by the development and refinement of production rules. Metacognitive proceduralization proposes a mechanism by which human cognition can become more skillful at monitoring and controlling its own states, such as attention, emotion, and, we suggest, metacognitive sensitivity.

Within this framework, metacognitive skill develops through three stages (Figure 1) similar to those of Fitts and Anderson, from an early stage of instruction following to an expert stage that relies on refined, automatic procedural knowledge (production rules).

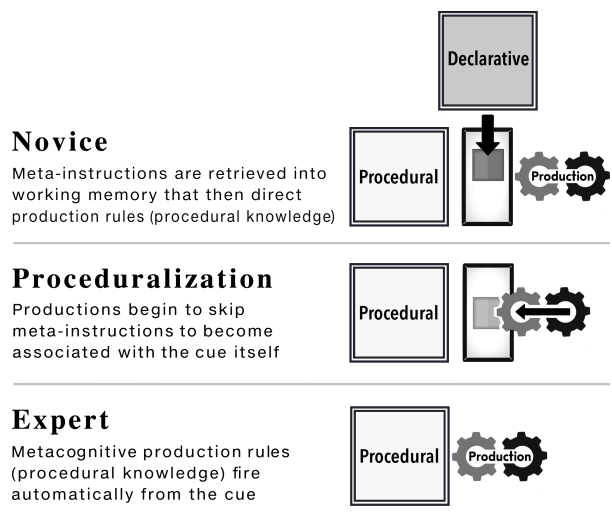

Figure 1: The three stages of metacognitive skill learning through proceduralization (Conway-Smith, West, & Mylopoulos, 2023).

The metacognitive practitioner progresses through the following three stages:

**The novice stage** begins with meta-instructions that direct monitoring and control resources in a specific way. In the case of metacognitive training in equanimity, meta-instructions direct the novice's attention toward the momentary fluctuations of affective experience (a feeling, sensation, or emotion). These meta-instructions are carried out by productions that retrieve them from declarative memory and execute them. Here, production speed-up could occur through mechanism 3 and possibly mechanism 1.

**The intermediate stage** of metacognitive training involves the process of proceduralization, where the practice of meta-instructions result in the creation of faster production rules to accomplish the task. Specifically, repeated practice would lead to the compilation of task-specific production rules that bypass declarative knowledge. Because they are faster (due to bypassing declarative memory and possibly being less complex), these productions are more strongly rewarded and more likely to bypass the retrieval of instructions in the future. Here, speed-up occurs through mechanism 2 and possibly mechanism 4 (with mechanism 3 and 1 still in play).

**The expert stage** involves a robust accumulation of production rules that have been refined and stored in procedural memory. These productions can be deployed automatically to act out monitoring and control processes quickly and effectively. Here, it is possible that productions accelerated through mechanisms 2 and 4 are so deeply engrained that fast productions resulting in metacognitive monitoring and control occur spontaneously. This would result in an increased ability to monitor, even without using mechanism 3 or 1 (although 3 and 1 would could still increase effectiveness if employed).

## Discussion

This paper investigates the empirical phenomenon where metacognitive training can effectively lower an individual's metacognitive threshold, thereby increasing perceptual access to their own internal cognitive states. To explore the underlying cognitive and computational processes of this phenomenon, we have employed the Common Model of Cognition with a special emphasis on the ACT-R framework. In the course of this investigation we have proposed a novel method of explanation by way of metacognitive proceduralization.